# Highly hydrogen-sensitive thermal desorption spectroscopy system for quantitative analysis of low hydrogen concentration (~$1 \times 10^{16}$ atoms/cm$^3$) in thin-film samples

Taku Hanna,[1] Hidenori Hiramatsu,[1,2,*] Isao Sakaguchi,[3] and Hideo Hosono[1,2]

[1] Materials Research Center for Element Strategy, Tokyo Institute of Technology, Mailbox SE-6, 4259 Nagatsuta-cho, Midori-ku, Yokohama 226-8503, Japan

[2] Laboratory for Materials and Structures, Institute of Innovative Research, Tokyo Institute of Technology, Mailbox R3-1, 4259 Nagatsuta-cho, Midori-ku, Yokohama 226-8503, Japan

[3] National Institute for Materials Science (NIMS), 1-1 Namiki, Tsukuba 305-0044, Japan

[*]Corresponding author
E-mail: h-hirama@mces.titech.ac.jp

Abstract

We developed a highly hydrogen-sensitive thermal desorption spectroscopy (HHS-TDS) system to detect and quantitatively analyze low hydrogen concentrations in thin films. The system was connected to an *in situ* sample-transfer chamber system, manipulators, and an rf magnetron sputtering thin-film deposition chamber under an ultra-high-vacuum (UHV) atmosphere of ~$10^{-8}$ Pa. The following key requirements were proposed in developing the HHS-TDS: (i) a low hydrogen residual partial pressure, (ii) a low hydrogen exhaust velocity, and (iii) minimization of hydrogen thermal desorption except from the bulk region of the thin films. To satisfy these requirements, appropriate materials and components were selected, and the system was constructed to extract the maximum performance from each component. Consequently, ~2000 times higher sensitivity to hydrogen than that of a commercially available UHV-TDS system was achieved using $H^+$-implanted Si samples. Quantitative analysis of an amorphous oxide semiconductor InGaZnO$_4$ thin film (1 cm × 1 cm × 1 μm thickness, hydrogen concentration of $4.5 \times 10^{17}$ atoms/cm$^3$) was demonstrated using the HHS-TDS system. This concentration level cannot be detected using UHV-TDS or secondary ion mass spectroscopy (SIMS) systems. The hydrogen detection limit of the HHS-TDS system was estimated to be ~$1 \times 10^{16}$ atoms/cm$^3$, which implies ~2 orders of magnitude higher sensitivity than that of SIMS and resonance nuclear reaction systems (~$10^{18}$ atoms/cm$^3$).

I. Introduction

Hydrogen is the most unavoidable and uncontrollable impurity in semiconductor devices because of the ease of contamination during fabrication processes, even under an ultra-high-vacuum (UHV) atmosphere. This impurity results in serious problems such as negative-bias-temperature instability in thin-film transistors [1, 2], unintentional carrier doping [3, 4], and the reduction of ferroelectric capacitors [5]. In addition to their effect on representative Si- and GaAs-based semiconductor devices [1, 6], the role of hydrogen impurities in oxide semiconductors such as ZnO has also attracted considerable attention. Even a low hydrogen concentration (~$10^{16}$ atoms/cm$^3$) in oxide semiconductors has been reported to result in unintentional $n$-type donors, thereby seriously affecting the electron transport properties and device performance [3]. To address these issues, highly hydrogen-sensitive (HHS) quantitative analyses of semiconductors have been performed using techniques including electron spin resonance [7], infrared spectroscopy [8, 9], Raman scattering spectroscopy [9], secondary ion mass spectrometry (SIMS) [10, 11], resonance nuclear reaction analysis (RNRA) [10, 12], and thermal desorption spectroscopy (TDS) [13 – 15].

The detection limit of these techniques strongly depends on the sample volumes because a larger total amount of hydrogen facilitates detection. Therefore, many quantitative hydrogen analyses have been performed on bulk samples with large volumes such as polycrystals and single crystals [9, 13]. However, semiconductor device technology is generally based on thin-film deposition and fine-scale processing technologies, for which additional material purification and refinement of processing techniques have been rapidly advancing. Therefore, the development of highly sensitive hydrogen detection techniques for small-volume samples such as thin films with nanometer-scale thicknesses is critical to investigate the effects of hydrogen impurities for practical semiconductor devices. However, the total amount of hydrogen in thin films is drastically less than that in bulk samples, leading to higher difficulty for highly sensitive hydrogen detection. Among the previously discussed hydrogen detection techniques, SIMS and RNRA are the most sensitive quantitative techniques for thin films (hydrogen detection limit ≈ $10^{18}$ atoms/cm$^3$) [11, 12]. However, quantitative analyses of extremely low hydrogen concentrations of < $10^{18}$ atoms/cm$^3$ have only been performed for bulk samples using optical detection [9] and TDS [13]. Thus, the development of hydrogen detection techniques for thin films with much higher sensitivity is strongly needed to better understand the role of hydrogen impurities and accurately control the hydrogen concentration in semiconductor devices.

As previously mentioned, TDS is a powerful technique for hydrogen detection. Unlike SIMS and RNRA, TDS has the advantage of providing information on the thermal stability and chemical states (i.e., chemical bonds and adsorption) as well as the concentration of desorbed hydrogen. Thermally desorbed hydrogen from samples is generally detected using a quadrupole mass spectrometer (QMS), i.e., a simple technique. Because the hydrogen detection limit for thin films of a commercially available TDS is ~$10^{19}$ atoms/cm$^3$ (i.e., one order of magnitude higher than that of SIMS and RNRA), a low hydrogen concentration in thin films (~$10^{18}$ atoms/cm$^3$), which was successfully determined using SIMS, could not be detected using a commercially available UHV-TDS system, as reported in [16]. Therefore, the development of a SIMS or RNRA system with higher hydrogen sensitivity would be more appropriate to achieve extremely high hydrogen-sensitive detection for thin films. However, for example, the beam spots of primary ions such as Cs$^+$ and Ar$^+$ of SIMS are focused to sizes of several tens of micrometers, indicating that the total amount of secondary hydrogen ions for detection is much smaller than that of TDS because all the hydrogen desorbed from the samples can be used as the detection target for TDS. Additionally, a large amount of funding, a long development period, and many labor sources are necessary for the development of a new SIMS or RNRA system. Therefore, we selected TDS, which has a simple and compact structure, for the development of highly hydrogen-sensitive detection technique. The development of such a system can be achieved with much less funding, a shorter development period, and several researchers by overcoming the issues in the present UHV-TDS system through effective design.

In this study, we developed a highly hydrogen-sensitive thermal desorption spectroscopy (HHS-TDS) system for quantitative analysis of low hydrogen concentrations in thin films. This article is categorized into the following sections. In Section II, we will briefly explain our design concept for the HHS-TDS. In Sections III–VII, we will present the actual approaches to development of the HHS-TDS system and its hydrogen detection performance. We successfully achieved approximately 2000 times higher hydrogen detection signals than those of a commercially available UHV-TDS system for a common $H^+$-implanted standard Si sample. The estimated hydrogen detection limit of the HHS-TDS system on an assumption is as sensitive as approximately $1 \times 10^{16}$ atoms/cm$^3$.

II. Key requirements for hydrogen detection by TDS: Design concept of the HHS-TDS

We first explain our design concept for the HHS-TDS system based on key requirements that we can propose from general equations on TDS.

Thermally desorbed hydrogen from a thin film on a substrate is generally detected with a QMS. The time-dependent change in the hydrogen partial pressure $P_{H2}(t)$, which is detected as the change in the hydrogen ion current with the QMS, is described by the mass-balance equation expressed in Eq. (1) [17] if ideal desorption conditions (i.e., no residual hydrogen gas in the TDS measurement chamber) are assumed:

$$V\left(\frac{dP_{H_2}(t)}{dt}\right) = -S\{P_{H_2}(t)\} + A\{q_{Sample}(t) + q'(t)\}, \qquad \text{Eq. (1)}$$

where $V$, $S$, $A$, $q_{sample}(t)$, and $q'(t)$ are the volume of the TDS measurement chamber, hydrogen exhaust velocity of the TDS system, surface area of the thin film on the substrate, hydrogen desorption rate from the bulk region of the thin film (this is the main target that we must measure precisely), and hydrogen desorption rate attributed to hydrogen-related impurities in the substrate and surface adsorption of hydrogen-related species such as water and hydrocarbons, respectively. When $S$ is much higher than the change rate of $P_{H2}(t)$ (i.e., $dP_{H2}(t)/dt$), we can approximate Eq. (1) using Eqs. (2) and (3) [18].

$$P_{H_2}(t) = \frac{A}{K_{300}S}\{q_{sample}(t) + q'(t)\} \qquad \text{Eq. (2)}$$

$$K_{300} = \frac{1}{RT} \qquad \text{Eq. (3)}$$

Here, $K_{300}$ is the molecule conversion factor at 300 K. In Eq. (3), $R$ and $T$ are the gas constant and temperature, respectively. Thus, $K_{300}$ is a constant in this case. Eq. (2) indicates that $P_{H2}(t)$ is proportional to the hydrogen desorption rate from the thin-film sample on the substrate (i.e., $q_{sample}(t)+q'(t)$) when $A$ and $S$ are constant (i.e., the same sample and TDS measurement system).

Based on Eq. (2), we can propose the following three key requirements for highly sensitive hydrogen detection by TDS:

(i) low $P_{H2}(t=0)$,

(ii) large $A$ and low $S$, and

(iii) low $q'(t)$.

For (i), a high signal-to-noise (S/N) ratio of hydrogen ion current in the QMS (i.e., high hydrogen sensitivity) is equivalent to a large difference between $P_{H2}(t)$ and $P_{H2}(t=0)$ during TDS measurements. Therefore, $P_{H2}(t=0)$ must be kept as low as possible (i.e., the residual hydrogen partial pressure in the TDS measurement chamber must be low). Thus, we designed a low-hydrogen-outgas and compact TDS system by selecting appropriate materials and components.

For (ii), although a larger $A$ is also effective for highly sensitive hydrogen detection (i.e., high $P_{H2}(t)$), we cannot treat large volume/area samples because the main target of this study is thin-film samples on substrates with low hydrogen concentrations and not large bulk samples such as polycrystals and single crystals. Thus, $S$ in the TDS measurement chamber must be reduced. Note that employing a high-$S$ vacuum exhaust system to achieve an UHV level has the opposite effect for highly sensitive hydrogen detection by TDS. To achieve low $S$, the application of a vacuum exhaust system with a high hydrogen compression ratio attained through low $S$ is important. Thus, we employed two turbomolecular pumps (TMPs) with high hydrogen compression ratios and constructed a tandem TMP system (i.e., a series-connecting 2 TMP system). In addition, an orifice was used to further reduce its $S$. The low residual hydrogen partial pressure discussed in (i) contributed to the UHV back pressure level of ~$10^{-10}$ Pa of the HHS-TDS measurement chamber irrespective of its low $S$.

For (iii), a minimum contribution of $q'(t)$ to $P_{H2}(t)$ is required because $q_{sample}(t)$ is the main target for TDS measurements. To satisfy this requirement, we constructed an *in situ* sample-transfer chamber system from a thin-film deposition chamber to the HHS-TDS measurement chamber without air exposure to avoid surface contamination before the TDS measurements. In addition, the substrates for thin-film deposition were thermally annealed under UHV atmosphere before deposition to minimize the hydrogen-related impurities in the substrate and surface adsorption species.

These concepts and the actual approaches used for the newly developed HHS-TDS system will be discussed in detail in Sections III–VI.

III. Set-up of the HHS-TDS system

Figures 1 and 2 present a schematic illustration and photographs of the HHS-TDS system with the *in situ* sample-transfer system and rf sputtering deposition chamber, respectively. Table I summarizes the important components of the HHS-TDS system. We will discuss the essential points for developing the HHS-TDS system in Sections III–VI following the order of components listed in Table I.

III-A. TDS measurement chamber

We considered the material and volume of the HHS-TDS measurement chamber to achieve a low residual hydrogen partial pressure. Hydrogen outgas from a vacuum chamber wall is one of the main origins of the high hydrogen residual pressure in general UHV chambers, which are usually composed of electropolished stainless steel, even if high temperature and long baking times are used. A high hydrogen outgas rate of $1 \times 10^{-12}$ Pa·m$^2$/s from an UHV chamber made of SUS 316 stainless was reported even after sufficient prebaking before/after assembly [19]. Therefore, we selected Be(0.2%)Cu (made by VacLab Inc., Japan), which is an ultralow hydrogen outgas material, as the material for the measurement chamber. In addition, all the metal parts in the HHS-TDS measurement chamber, including Be(0.2%)Cu, were thermally annealed at 400 °C for 72 h under an UHV atmosphere before being assembled. Consequently, a low hydrogen outgas rate of $5.6 \times 10^{-14}$ Pa·m$^2$/s was achieved [19], which is more than two orders of magnitude lower than that of the previously discussed stainless steel case.

The small volume of the HHS-TDS measurement chamber (565 ml), which is much smaller than that of commercially available UHV-TDS systems (usually several liters), also contributed to the low hydrogen outgas rate because of the small surface area of the measurement chamber wall.

III-B. Vacuum exhaust system

The exhaust system of the HHS-TDS measurement chamber was designed to achieve an excellent UHV back pressure level of $9 \times 10^{-10}$ Pa without employing a high-exhaust-velocity TMP (see Eq. (2); achieving low $S$ is important for highly hydrogen-sensitive TDS.). Because the HHS-TDS measurement chamber was composed of low-hydrogen-outgas components, a high hydrogen gas compression ratio of the exhaust system was more critical than a high exhaust velocity to decrease the amount of residual hydrogen gas in the measurement chamber. To achieve this high hydrogen gas compression ratio, we employed a series-connecting 2 TMP system with high hydrogen compression ratios (called a tandem TMP structure; see Table I for further details). In addition, the as-yet-high exhaust velocity was reduced to 1 L/s for N$_2$ and 3.74 L/s for $H_2$ with a 3.3-mm-diameter Cu orifice on top of the tandem TMP #1. The orifice effectively suppressed the counter flow of hydrogen gas from the tandem TMP#1, leading to further improvement of the hydrogen sensitivity of the HHS-TDS system.

Commercially available general vacuum gauges cannot detect UHV vacuum levels of $\sim 10^{-10}$ Pa because they cannot accurately detect small amounts of hydrogen desorption because of their self-hydrogen-outgas mainly from an ion source. Therefore, a specially remodeled ultralow outgas vacuum gauge (3B Gauge [20], VacLab Inc., Japan), in which a hot cathode ion source was housed on a Be(0.2%)Cu flange, was employed. We calibrated sensitivity factors for $N_2$, $H_2$, and $D_2$ gases of the 3B Gauge using a spinning rotor gauge.

For further improvement of the hydrogen residual partial pressure, we employed an additional vacuum exhaust pump, a non-evaporable getter (NEG) pump (Magic NEG, VacLab Inc., Japan) separated with an all-metal gate valve from the HHS-TDS measurement chamber. The vacuum level of the Magic NEG when the effective exhaust velocity for $H_2$ = 0 L/s is $\sim 10^{-10}$ Pa, which is a one-order-of-magnitude-higher UHV level than that of a general NEG pump ($\sim 10^{-9}$ Pa), mainly because the housing material is Be(0.2%)Cu. This indicates that the Magic NEG can effectively work at an UHV level of $\sim 10^{-10}$ Pa even though general NEG pumps do not work well at this high UHV level. The Magic NEG was used directly before the TDS measurements to further reduce the residual hydrogen partial pressure.

The reduced exhaust velocity with the orifice is sufficiently higher than the hydrogen outgas rate during the HHS-TDS measurements. Even though the exhaust velocity is low, the vacuum level in the measurement chamber can be kept below $5 \times 10^{-9}$ Pa if the QMS works.

III-C. QMS

The hydrogen ion current signal during TDS measurements, measured with commercially available QMS equipped with a Faraday cup, is generally amplified to $10^4$–$10^5$ times higher levels with a secondary electron multiplier (SEM) because of its generally poor S/N ratio for hydrogen detection. However, the amplification factor of the SEM for hydrogen typically fluctuates with changes in measurement conditions such as time and temperature even if a pre-calibrated constant amplification voltage is applied. This fluctuation has a negative effect on the reliability of hydrogen detection with a TDS system, especially for the quantitative analysis of low hydrogen concentrations in thin-film samples. Therefore, we employed an ultralow outgas QMS (WAT-Mass [21, 22], VacLab Inc., Japan). An ion source housed on a Be(0.2%)Cu flange such as 3B Gauge (see Section III-ii) was employed, leading to a much lower hydrogen outgas rate of $3.4 \times 10^{-12}$ Pa·m$^3$/s than that of a commercially available QMS ($7.5 \times 10^{-8}$ Pa·m$^3$/s) [22]. The hydrogen ion current of the WAT-Mass was directly measured using only a Faraday cup without SEM amplification because the S/N ratio of the hydrogen ion current is drastically improved by the high hydrogen sensitivity and low hydrogen residual pressure of the HHS-TDS system. This direct (i.e., not amplified) measurement of the hydrogen ion current results in good linearity of the hydrogen and deuterium signal calibration, free from the harmful effect of SEM amplification, such as encountered for general QMS analysis. For the QMS calibration results, see Section IV-B.

The mounting positions of the vacuum gauge (3B Gauge) and QMS (WAT-Mass) are also important factors for the precise measurement of the thermally desorbed hydrogen partial pressure (see Figs. 1 and 2(b)). These instruments are mounted at equivalent positions in the measurement chamber against the hydrogen exhaust gas flow line (from the sample to the tandem TMP); i.e., they are mounted at face-to-face positions where both positions are equivalent from the viewpoint of exhausted hydrogen gas.

III-D. Sample heating system and sample stage

Thermal conduction from heated samples during TDS measurements is a serious issue because it causes unpredictable hydrogen outgas, which leads to an unreliable low hydrogen detection limit (i.e., an increase in hydrogen residual pressures during TDS measurements) and incorrect quantitative analysis. The absolute value of hydrogen-desorbed temperatures can be a good gauge of the hydrogen chemical state in samples. From this viewpoint, precise temperature calibration is critical for TDS.

A lamp heating system is typically employed in commercially available UHV-TDS systems to heat the sample as a contactless heating method is desired to suppress the hydrogen outgas as much as possible. Because UHV-TDS measurement chambers are generally equipped with a lamp controlled by a thermocouple, thermal conduction is unavoidable. Additionally, the sample is typically heated on a face-contact silica-glass stage, which leads to high thermal conduction of UHV-TDS systems because of the large contact areas.

Thus, we developed an entirely different contactless heating system from outside of the HHS-TDS measurement chamber using an infrared laser diode (LU0915C300, Lumics GmbH, wavelength = 915 nm, maximum power = 300 W) with a collimated aspherical lens to heat the samples to 1000 °C. A customized point-contact high-purity silica-glass stage ($H_2$ impurity: $\leq 5\times10^{15}$ molecules/cm$^3$, Shin-Etsu Chemical Co.) was fabricated to minimize thermal conduction from the heated sample to the HHS-TDS measurement chamber. Only four edges of the sample (lateral size: 1 cm × 1 cm) contacted the stage to achieve low thermal conductivity during the TDS measurements (see Fig. 2). The rough, sandblast-treated-like (i.e., not transparent) surface of the silica-glass stage is observed in the bottom photograph in Fig. 2(c). This rough surface that was unintentionally created during the fabrication process is also effective because of its low thermal conduction.

The sample temperature ($T_{sample}$) during TDS measurements is generally controlled with a thermocouple directly attached to the heated sample and/or sample stage; however, this measurement causes unpredictable hydrogen outgas because both the heated thermocouple and corresponding thermal conduction from the thermocouple can be serious hydrogen outgas sources. To avoid such harmful hydrogen outgas, $T_{sample}$ of the HHS-TDS system is controlled only by the dc input current for the laser diode using a high-increment dc current source (PAN110-3A, Kikusui Electronics Co., minimum current increment: 1 mA, maximum output current: 3 A) without a thermocouple in the HHS-TDS measurement chamber. Using this approach, we succeeded in removing both the heating source and temperature measurement equipment from inside of the HHS-TDS measurement chamber. However, this technique requires the precise calibration of $T_{sample}$ because $T_{sample}$ is controlled only by the programmed dc input current to the laser diode without *in situ* or *real-time* $T_{sample}$ monitoring and feedback using a thermocouple. Temperature calibration will be discussed in Section IV.

IV. Calibration of $T_{sample}$ and QMS

IV-A. $T_{sample}$ calibration

Figure 3 presents the $T_{sample}$ calibration measurements for the HHS-TDS system employing the laser diode heating system. $T_{sample}$ was calibrated with a K-type thermocouple (chromel/alumel) directly connected to the bottom of a Si wafer substrate (see the inset of Fig. 3(a) for a schematic illustration of the calibration experiment setup). This setup is used only when $T_{sample}$ calibration is performed; the thermocouple connected to the Si is then removed from the HHS-TDS measurement chamber to minimize hydrogen outgas.

In Fig. 3(a), $T_{sample}$ in the HHS-TDS system is plotted as a function of the dc input current to the laser diode heating system. To precisely calibrate $T_{sample}$, we allowed sufficient saturation (> 5 min) of the laser irradiation at each $T_{sample}$ data point. This relationship between $T_{sample}$ and the dc input current was used as the calibration curve of $T_{sample}$ for all the HHS-TDS measurements in this study.

Figure 3(b) shows the relationship between the calibrated $T_{sample}$ and measurement time at a heating rate of 0.55 °C/s ($T_{sequence}$), indicating that the heat sequence employed in this study showed good agreement with the calibrated $T_{sample}$. However, a slightly slow response to $T_{sequence}$ was observed in the low-temperature region from room temperature to ~300 °C. In this temperature range, the operating mode of the laser diode drastically changes from the diode to lasing mode, as observed near the input current of 0.5 A (i.e., the threshold current of the laser diode) in Fig. 3(a). However, we employed this heating rate of 0.55 °C/s because slower and faster $T_{sequence}$ lead to low hydrogen sensitivity and poor temperature response, respectively.

IV-B. Calibration of QMS for hydrogen quantitative analysis

Equation 4 expresses the relationship between the hydrogen ion current $I_{H2}$ and partial pressure $P_{H2}$ detected with a QMS if SEM amplification is not applied (i.e., only a Faraday cup is used) [20].

$$I_{H_2} = FP_{H_2}I_e \qquad \text{Eq. (4)}$$

Here, $F$ and $I_e$ are a sensitivity factor for hydrogen and the emission current, respectively. Because both parameters are generally constant, a proportional relationship between $I_{H2}$ and $P_{H2}$ is observed under ideal desorption conditions (i.e., no residual hydrogen gas in the TDS measurement chamber.). To examine the reliability of hydrogen detection with the QMS of the HHS-TDS (WAT-Mass), the relationship between the ion current measured with the QMS and the partial pressure of externally introduced hydrogen ($m/z = 2$) and deuterium ($m/z = 4$) gases measured by the vacuum gauge (3B gauge) was investigated (see Fig. 4.). Hydrogen gas (purity: > 99.99999 vol.% (Grade 1)) or deuterium gas (purity: 99.6 vol.%) was precisely introduced from a gas-introduction port through a variable leak valve to the HHS-TDS measurement chamber, which was initially placed under vacuum at $5 \times 10^{-9}$ Pa for each measurement. Then, the QMS ion currents and partial pressures were measured. A completely proportional relationship was observed with slopes of 1 (±0.009) for hydrogen and 1 (±0.008) for deuterium, which precisely follows Eq. (4), in a wide partial pressure range up to $1 \times 10^{-2}$ Pa. This result indicates that the HHS-TDS system can accurately measure changes in hydrogen and deuterium partial pressure within approximately 1% error under an UHV back pressure of $\sim 5 \times 10^{-9}$ Pa without effects of residual hydrogen gas.

For quantitative hydrogen analysis based on a TDS spectrum, the total amount of thermally desorbed hydrogen $Q$ is described by Eq. (5), which was derived from a time integration of Eq. (2) [17],

$$Q = \int_{t_1}^{t_2} q(t)\, dt = \left(\frac{SK_{300}}{A}\right) \int_{t1}^{t2} \left(\frac{dP_{H_2}}{dt}\right) dt. \qquad \text{Eq. (5)}$$

Here, $V$, $S$, $K_{300}$, and $A$ are constant. Therefore, Eq. (5) indicates the existence of a proportional relationship between $Q$ and the integration area of $P_{H2}$ under ideal desorption conditions (i.e., no residual hydrogen gas in the TDS measurement chamber).

Based on Eq. (5), we performed calibration experiments for exact hydrogen quantitative analysis by thermal desorption using hydrogen- and deuterium-ion-implanted standard Si samples. An ion-implantation apparatus (RD-200L, Nissin Ion Equipment Co. Ltd., acceleration energy: 45 keV) at the National Institute for Materials Science (NIMS, Japan) was employed for fabrication of all the standard ion-implanted samples discussed in this section. Because air exposure of the implanted standard samples was unavoidable (i.e., no *in situ* connection to the HHS-TDS), hydrogen-implanted Si standard samples with relatively high dose concentrations (total amount of implanted hydrogen ions = $1.3 \times 10^{15} - 3.7 \times 10^{16}$ ions, calculated as dose amounts (ions/cm$^2$) × dosed area (cm$^2$)) were selected. The effect of hydrogen-related impurities in the Si substrate and surface adsorption of hydrogen-related species such as water and hydrocarbons could not be eliminated because of the unavoidable air exposure of the samples (i.e., the effect of the $q\cdot(t)$ term in Eq. (2)). These effects will be discussed in Section VI and Fig. 8. Therefore, we selected deuterium as another standard implantation ion for the lower implantation region of $4.9 \times 10^{12} - 5.0 \times 10^{14}$ ions because deuterium is one of the isotopes of hydrogen and its relative isotopic abundance is as low as approximately 0.014%.

Figure 5 shows the relationship between the integrated area of the $H^+$- or $D^+$-ion current and the total numbers of $H^+$- or $D^+$-implanted ions (see Fig. S1 for the raw spectra). All the data for hydrogen were obtained after subtraction of the baseline signal only from the Si substrate. $T_{sequence}$ at the heating rate of 0.55 °C/s was employed up to 1000 °C. As expected, a proportional relationship following Eq. (5) was observed for both the thermally desorbed hydrogen and deuterium from the ion-implanted standard samples in the wide range of $4.9 \times 10^{12}$–$3.7 \times 10^{16}$ ions. Each slope is very close to 1 (results of least-squares fittings: 1.00 (±0.1) for hydrogen and 0.99 (± 0.07) for deuterium). We confirmed a linear relationship with a slope of 1 for the QMS integrated ion currents of both $H_2$ and $D_2$ gases, as shown in Fig. 4. Thus, we judged that the hydrogen calibration line can be linearly extrapolated to the lower dose concentration region, as shown in Fig. 5, even though only the highly dosed standard samples were measured for desorbed hydrogen.

These results indicate that the HHS-TDS system can also quantitatively detect thermally desorbed $H_2$ and $D_2$ (and not only $H_2$ and $D_2$ gas sources) with a completely linear relationship over a wide range. We employed this calibration line for hydrogen in Fig. 5 for the highly sensitive hydrogen quantitative analysis using the HHS-TDS system in this study.

The maximum S/N ratio of the HHS-TDS spectra was $2.7 \times 10^4$ in the $3.7 \times 10^{16}$ ion hydrogen-implanted standard sample (see Fig. S1) with a maximum hydrogen partial pressure of $4.5 \times 10^{-6}$ Pa. Assuming that the detection limit is an S/N ratio of 3 (i.e., approximately $1/10^4$ of the present S/N ratio), we can roughly estimate the hydrogen detection limit of the HHS-TDS system to be a hydrogen desorption rate of $1.9 \times 10^{-12}$ Pa·m$^3$/s [= $3/(2.7\times10^4)$ × (hydrogen partial pressure) × (hydrogen exhaust velocity = $3.74\times10^{-3}$ m$^3$/s, see Table I)]. This detection limit is slightly better than the reported value ($3.4 \times 10^{-12}$ Pa·m$^3$/s) for the same-type QMS [22], implying that the hydrogen detection performance of the ultralow-outgas QMS employed in this study is not disturbed by the other components in the HHS-TDS system and that we succeeded in constructing a good low-outgas hydrogen measurement environment.

V. Comparison of the HHS-TDS with UHV-TDS

The hydrogen sensitivity of the HHS-TDS system was compared with that of a commercially available UHV-TDS system (TDS1400TV, ESCO Ltd., Japan). As discussed in Section VI and Fig. 8, we cannot eliminate the harmful effect to the hydrogen ion current originating from the hydrogen-related impurities in the Si substrate and surface adsorption of hydrogen-related species such as water and hydrocarbons due to air exposure of the samples. To minimize this effect, we employed a commercially available Si standard sample (1 cm × 1 cm × 500 μm in thickness) with a relatively high H$^+$ dose concentration (dose amount: $1 \times 10^{17}$ ions/cm$^2$). Because the harmful effect relatively decreases due to the high dose concentration, the effect can almost be eliminated for a high-dose standard sample. Then, we estimated the hydrogen detection limit from the S/N ratios of the obtained spectra.

Figure 6 presents the TDS spectra for $m/z$ = 2 of the common H$^+$-implanted Si standard samples (H$^+$ = $1 \times 10^{17}$ ions/cm$^2$) obtained using two types of TDS. The intensity of the unamplified ion current for the HHS-TDS system and of the amplified ion current for the UHV-TDS system were normalized as the baseline signals at room temperature (measurement time = 0 s). The observed maximum S/N ratios of the HHS-TDS and UHV-TDS spectra were $1.2 \times 10^5$ and $5.1 \times 10^1$, respectively. A gradual increase in the baseline with increasing temperature ($T_{sample} > 400$ °C) is observed in the HHS-TDS spectrum, which mainly originates from hydrogen outgas from the Si substrate (data can be shown in Fig. 8). Note that the hydrogen sensitivity of the HHS-TDS system is approximately 2000 times higher than that of the commercially available UHV-TDS system even without amplification. This high sensitivity (i.e., both of high S/N ratio and high signal absolute intensity) originates from low hydrogen residual pressure in the HHS-TDS measurement chamber with the low exhaust velocity using the orifice. Assuming that the detection limit is a S/N ratio of 3 and that the integrated areas of the HHS-TDS spectra are proportional to the maximum $H_2$ ion current intensity, the hydrogen detection limits were estimated to be $2.6 \times 10^{16}$ and $5.9 \times 10^{19}$ atoms/cm$^3$ for the HHS-TDS and UHV-TDS systems, respectively.

VI. *In situ* sample-transfer chamber system with manipulators and pre-treatment of Si substrate before thin-film deposition: Hydrogen impurity concentration of Si substrate

As discussed in Section II, a low $q'(t)$ is one of the key requirements for highly sensitive hydrogen detection using TDS. To experimentally examine this parameter, we developed an *in situ* sample-transfer chamber system to transfer the sample without exposure to air and manipulators to pick the sample up from a substrate carrier and place it on the HHS-TDS sample stage under an UHV atmosphere. The *in situ* sample-transfer chamber (see Fig. 1 and 2) is connected to an rf magnetron sputtering thin-film deposition chamber. This transfer chamber is placed under vacuum using ion pumps (Vaclon Plus 500 Combination Pump, Agilent Technologies, Inc., exhaust velocities: 880 L/s for $N_2$, 1930 L/s for $H_2$) with Ti sublimation pumps and liquid nitrogen shrouds to maintain a vacuum level of ~$10^{-8}$ Pa. A transfer carrier in this system can carry a maximum of three substrate carriers from the deposition chamber just near the HHS-TDS measurement chamber (see Fig. 7(a)). There are two types of manipulators between the HHS-TDS measurement chamber and *in situ* transfer chamber system. One is a sample manipulator to pick the sample up from the substrate carrier in the *in situ* transfer chamber, and the other is a sample stage manipulator for the sample stage in the HHS-TDS measurement chamber. Figure 7 shows the *in situ* sample transfer technique using these systems (see caption in FIG. 7 for each transfer action in detail.). By employing these systems, we succeeded in completely avoiding air exposure of the deposited thin-film samples for the HHS-TDS measurements.

Next, we measured HHS-TDS spectra for $m/z = 2$ for two types of Si substrates (purity: 11 N) to examine the effect of the hydrogen-related species in the Si substrate and surface adsorption of hydrogen-related species such as water and hydrocarbons due to air exposure of the samples. One of the substrates was an as-received sample that was transferred from air to the HHS-TDS measurement chamber without any pre-treatment. The other substrate was first transferred from air to the rf sputtering deposition chamber and then thermally annealed at 800 °C for 1 h in the deposition chamber (This temperature is the maximum value for our deposition chamber.). The annealed substrate was then transferred *in situ* to the HHS-TDS measurement chamber after being stored in the deposition chamber under an UHV atmosphere for 1 day without exposure to air.

Figure 8 presents the HHS-TDS spectra for $m/z = 2$ for the two types of Si substrates. The maximum ion current for the as-received sample was 1000 times higher than that of the pre-treated and *in situ* transferred sample. The hydrogen species concentrations in the as-received and annealed and *in situ* transferred Si substrates were $2.4 \times 10^{17}$ and $4.5 \times 10^{14}$ atoms/cm$^3$, respectively. These results indicate that large amounts of surface-adsorbed and impurity-hydrogen species in the Si wafer were detected when it was transferred from the air. Note that a hydrogen concentration of $10^{17}$ atoms/cm$^3$ is of the same order as that in a thin-film oxide semiconductor, which is discussed in Section VII. However, these impurities are almost completely removed to the background level of the QMS ($5 \times 10^{-14}$ A) by employing the pre-annealing step and *in situ* transfer.

Such uncontrollable hydrogen desorption except from the bulk region of the samples is not generally considered for UHV-TDS measurements because the hydrogen desorption rates from large bulk samples and high-hydrogen-concentration thin-film samples are significantly higher than those of other hydrogen outgas sources. However, detection of extremely small amounts of hydrogen desorption from thin-film samples with low hydrogen concentrations requires the elimination of such undesirable hydrogen desorption. Therefore, the combination of HHS-TDS and this *in situ* sample transfer system is a powerful method to quantitatively detect intrinsically included low hydrogen concentrations by providing a low $q(t)$ measurement condition.

VII. Quantitative analysis of hydrogen concentration in a thin-film oxide semiconductor using the HHS-TDS without air exposure

We demonstrated the hydrogen quantitative analysis of a thin-film sample with a size of 1 cm × 1 cm × 1 μm in thickness. The film was a representative thin-film amorphous oxide semiconductor, InGaZnO$_4$ (*a*-IGZO) [23, 24], which is practically applied in thin-film transistors for flat-panel displays instead of *a*-Si. A multitarget rf magnetron sputtering deposition chamber (Eiko Co., Japan) with an UHV back pressure of ~$10^{-8}$

Pa was employed for film deposition to minimize the hydrogen impurity concentration in the *a*-IGZO thin films because hydrogen concentration in the *a*-IGZO thin films largely depends on the base pressure of the sputtering deposition chamber [16]. Before deposition, the Si substrates were pre-annealed at 800 °C for 1 h in the deposition chamber, as described in Section VI and Fig. 8. Then, Ar (purity: > 99.9999 vol.%, Grade 1) and $O_2$ (> 99.99995 vol.%, Grade 1) mixed gas with a total pressure of 0.3 Pa was introduced into the deposition chamber (Ar:$O_2$ flow ratio = 19.8:0.2 in standard cc per minute). The *a*-IGZO thin films were deposited on the annealed thin Si substrates at room temperature. Three 2-inch ceramic sputtering targets (purity: $In_2O_3$ = 99.9 %, $Ga_2O_3$ = 99.99 %, and ZnO = 99.99 %) were used, and the rf power of each of the cathodes was independently optimized. The chemical composition ratio (In:Ga:Zn) of the obtained *a*-IGZO films measured by X-ray fluorescence spectroscopy was 1.06:1.01:0.95 (i.e., very close to the cation composition ratio in bulk $InGaZnO_4$). The optical band gap was estimated to be approximately 3 eV with low subgap-state absorption (see Fig. S2 for the optical absorption spectrum). These composition and optical results are similar to those of previously reported *a*-IGZO thin films fabricated using an UHV sputtering chamber [16, 25].

Figure 9 presents the HHS-TDS spectra for *m*/*z* = 2 of the *a*-IGZO thin film and pre-annealed Si substrate. Clear ion current signals originating from thermally desorbed hydrogen were observed in the HHS-TDS spectrum of the *a*-IGZO thin film. The ion current signal from the *a*-IGZO thin film was sufficiently high compared with that of the pre-annealed Si substrate. The hydrogen concentration in the *a*-IGZO thin film was determined to be $4.5 \times 10^{17}$ atoms/cm$^3$, which is a low hydrogen concentration that cannot be detected quantitatively by an UHV-TDS system, based on the calibration in Fig. 5 and considering the sample size (i.e., 1 cm × 1 cm × 1 μm in thickness). Note that the hydrogen concentration in the *a*-IGZO thin film is of the same order as that in the as-received Si substrate ($2.4 \times 10^{17}$ atoms/cm$^3$, see Fig. 8). Therefore, pre-thermal treatment of substrates to minimize the effect of impurities and adsorbed hydrogen is necessary for quantitative detection of low hydrogen concentrations.

For comparison, an *a*-IGZO thin film deposited under the same conditions was measured using SIMS (CAMECA, IMS-4f, Primary ion: Cs$^+$), which is another quantitative analysis technique with high hydrogen sensitivity. To reduce surface desorption of hydrogen-related species, surface sputtering was performed for 5 min before the SIMS measurement. As observed in the SIMS depth profile (see the inset of Fig. 9), the level of the *m/z* = 1 signal originating from H$^-$ was equal to that of the background hydrogen signal in the SIMS measurement chamber, indicating that the hydrogen signal of $4.5 \times 10^{17}$ atoms/cm$^3$ was not detected because of the detection limit of $6.6 \times 10^{18}$ atoms/cm$^3$ in the present analysis.

The obtained maximum S/N ratio of the hydrogen signal from the *a*-IGZO thin film was $1.4 \times 10^2$. Assuming that the detection limit is a S/N ratio of 3 and that the integrated area of the HHS-TDS spectra is proportional to the maximum H$_2$ ion current intensity, the estimated hydrogen detection limit of the HHS-TDS system is as sensitive as $9.5 \times 10^{15}$ atoms/cm$^3$, which is ~2 orders of magnitude higher than that of SIMS and RNRA, within a 10% error.

VIII. Summary

A highly hydrogen-sensitive thermal desorption spectroscopy (HHS-TDS) system including an *in situ* sample transfer chamber, manipulators, and a rf magnetron sputtering thin-film deposition chamber was developed to detect and quantitatively analyze low hydrogen signals from thin-film samples (i.e., signal from small sample volumes). The present work is summarized as follows:

(1) We proposed three key requirements to achieve high hydrogen sensitivity using this system: (i) a low hydrogen residual partial pressure, (ii) a low hydrogen exhaust velocity, and (iii) minimum contribution of hydrogen thermal desorption except from the bulk region of thin-film samples. To satisfy these requirements, we selected appropriate materials and components and constructed the system to extract the maximum performance of each component (FIGs. 1, 2 & 7 and Table I). The sample temperature, which was controlled by a laser diode heating system, and the sensitivity of the ultralow outgas QMS were calibrated to precisely measure the desorption temperature and hydrogen ion current for quantitative analysis, respectively (FIGs. 3 – 5).

(2) Two types of HHS-TDS measurements were performed to estimate the hydrogen detection limit. One measurement compared the hydrogen detection performance of the HHS-TDS system and a commercially available UHV-TDS system using a common commercially available $H^+$-implanted Si standard sample (size: 1 cm × 1 cm × 500 μm in thickness) with a relatively high $H^+$-dose concentration ($H^+ = 1 \times 10^{17}$ ions/cm$^2$, see FIG. 6). We achieved approximately 2000 times higher hydrogen detection signals for the HHS-TDS system than for the commercially available UHV-TDS system. The other measurement was for a thin-film amorphous oxide semiconductor, InGaZnO$_4$ (*a*-IGZO), with a size of 1 cm × 1 cm × 1 μm in thickness on a Si substrate, which was pre-thermally annealed at 800 °C and then *in situ* transferred to the film deposition chamber (see FIG. 9).

(3) The pre-thermal annealing treatment of the Si substrates with *in situ* transfer was effective for highly sensitive hydrogen detection because the effect of hydrogen-related impurities in the Si substrate and surface adsorption of hydrogen-related species is serious when highly sensitive quantitative analysis is performed, especially for thin-film samples with low hydrogen concentrations. The HHS-TDS spectra for the Si substrates revealed large hydrogen concentrations ($2.4 \times 10^{17}$ atoms/cm$^3$), which disturb highly sensitive hydrogen experiments, when the Si substrate was not pre-treated (i.e., as received, see FIG. 8).

(4) The hydrogen concentration in an *a*-IGZO thin-film sample with a size of 1 cm × 1 cm × 1 μm (thickness) was quantitatively determined to be $4.5 \times 10^{17}$ atoms/cm$^3$; this concentration cannot be quantitatively detected using UHV-TDS and SIMS systems.

(5) The estimated hydrogen detection limit of the HHS-TDS system is ~$1 \times 10^{16}$ atoms/cm$^3$ [$2.6 \times 10^{16}$ atoms/cm$^3$ (estimated from the former demonstration) and $9.5 \times 10^{15}$ atoms/cm$^3$ (from the latter one)], which is ~2 orders of magnitude higher than that of SIMS and RNRA based on the assumptions that the detection limit is a S/N ratio of 3 and that the integrated area of the HHS-TDS spectrum is proportional to the maximum H$_2$ ion current intensity.

The HHS-TDS system developed in this work is expected to contribute to high-sensitivity hydrogen detection and quantitative low-concentration analysis for semiconductor and ferroelectric thin-film devices. The performance of these devices is thought to be affected by the presence of small hydrogen concentrations, which have thus far been masked because of the lack of such highly hydrogen-sensitive and user-friendly techniques.

Supplementary Material

See supplementary material for the HHS-TDS spectra of $H^+$- and $D^+$-implanted standard Si samples and an optical absorption coefficient spectrum of an *a*-IGZO thin film on a silica-glass substrate.

Acknowledgments

This work was supported by the Ministry of Education, Culture, Sports, Science, and Technology (MEXT) through the Element Strategy Initiative to Form Core Research Center. T. H. would like to acknowledge Dr. Fumio Watanabe and Mr. Reiki Watanabe (VacLab Inc., Japan) for fruitful discussions on vacuum technology. H. Hi. acknowledges Mr. Makoto Ouchi (Eiko Co., Japan) for valuable discussions on the design of the *in situ* sample transfer system and manipulators. H. Hi. was also supported by the Japan Society for the Promotion of Science through a Grant-in-Aid for Scientific Research on Innovative Areas "Nano Informatics" (Grant Number 25106007) and Support for Tokyotech Advanced Research (STAR).

TABLE I. Components of the HHS-TDS system with the *in situ* transfer system. The components of commercially available UHV-TDS systems are also provided for comparison.

Footnotes:

a: Data taken as examples from refs [19, 20, 22].

b: Typical data of an UHV-TDS (TDS1400TV, ESCO, Japan) at our laboratory.

c: The base pressure of the HHS-TDS measurement chamber is $< 5\times10^{-9}$ Pa when the ultralow outgas QMS (WAT-Mass, VacLab Inc., Japan) works.

d: TMP = turbomolecular pump. A series-connecting 2 TMP system is called a tandem TMP.

e: Here, two ICF70-type NEG pumps are compared.

f: The UHV-TDS (TDS1400TV) at our laboratory employs a QMS (PrismaPlus QMG 220 F2, Pfeiffer Vacuum GmbH).

g: The sample temperature is calibrated before the HHS-TDS measurements with a K-type thermocouple directly connected to the bottom of a Si wafer substrate (see Fig. 3).

| Component | | HHS-TDS (this work) | UHV-TDS |
|---|---|---|---|
| TDS measurement chamber | Material | Be(0.2%)Cu (made by VacLab Inc.) | Electro-polished stainless steel (made of SUS316) |
| | Outgas rate [19] | $5.6\times10^{-14}$ Pa·m$^2$/s (Pre-baked at 400°C for 72h under UHV atmosphere) | $1\times10^{-12}$ Pa·m$^2$/s (Pre-baked at 960°C for 25h under UHV atmosphere) [a] |
| | Volume | 565 ml | Several liters (ex. 7 L [b]) |
| | Base pressure | $9\times10^{-10}$ Pa [c] | ca. $5\times10^{-8}$ Pa [b] |

| | Method | Tandem TMP [d] | | Single TMP |
|---|---|---|---|---|
| Vacuum exhaust system for TDS measurement chamber | Product name | For tandem #1: Hipace300 (made by (Pfeiffer Vacuum GmbH) | For tandem #2: TG70F (made by Osaka Vacuum Ltd.) | MAG W400 (made by Oerlikon Leybold Vacuum GmbH) [b] |
| | Compression ratio for $H_2$ | $> 9 \times 10^5$ | $> 1 \times 10^5$ | $3.2 \times 10^5$ |
| | Exhaust velocity | $N_2$: 260 L/s $H_2$: 220 L/s | $N_2$: 70 L/s $H_2$: 45 L/s | $N_2$: 365 L/s $H_2$: 200 L/s |
| | Remark | Exhaust velocity is finally reduced with an orifice at the top of tandem TMP #1 into 1 L/s (for $N_2$) and 3.74 L/s (for $H_2$) | | – |
| Additional vacuum exhaust pump | Type | Non-evaporable getter (NEG) pump | | Not equipped. (The following is information on a general NEG pump.) |
| | Product name [e] | Magic NEG (made by VacLab Inc.) | | NEXTorr D 100-5 (made by SAES Getters) |
| | Housing material | Be(0.2%)Cu | | Not specified |
| | Exhaust velocity | $N_2$: 25 L/s, $H_2$: 70 L/s | | $N_2$: 40 L/s, $H_2$: 100 L/s |
| | Vacuum level at effective exhaust velocity for $H_2$ = 0 L/s | ca. $10^{-10}$ Pa | | ca. $10^{-9}$ Pa |

| | | | |
|---|---|---|---|
| Vacuum gauge | Product name | 3B Gauge (made by VacLab Inc.) | General BA gauge |
| | Feature for low $H_2$ outgas rate | Ion source housed on Be(0.2%)Cu flange | Not specified |
| | Detection limit [20] | $5.4 \times 10^{-12}$ Pa | ca. $1 \times 10^{-9}$ Pa [a] |
| Quadrupole mass spectrometer (QMS) | Product name | WAT-Mass (made by VacLab Inc.) | Transpector H100M (made by Leybold Inficon Inc.) [a, f] |
| | Feature for low $H_2$ outgas rate | Ion source housed on Be(0.2%)Cu flange | Not specified |
| | Outgas rate for $H_2$ [22] | $3.4 \times 10^{-12}$ Pa·m$^3$/s | $7.5 \times 10^{-8}$ Pa·m$^3$/s [a] |
| | Detection method | Faraday cup (not amplified in this work) | Amplified by secondary electron multiplier (SEM) |
| Sample heating | Source | Infrared laser diode | Lamp heater |
| | Method | Irradiation from outside of TDS measurement chamber | Heating inside TDS measurement chamber |
| | Product name | LU0915C300 (made by Lumics GmbH) | – |
| | Temperature control | DC input current to laser diode [g] | Thermocouple inside TDS measurement chamber |
| | Remark | Wavelength: 915 nm, maximum power: 300 W | – |
| Sample stage | Material | High-purity silica-glass (made by Shin-Etsu Chemical Co.) | General silica-glass |
| | $H_2$ impurity | ≤ $5 \times 10^{15}$ molecules/cm$^3$ | – |
| | Contact between sample and TDS stage | Point-contact at 4-corner edges of sample | Face contact |

| Sample transfer | An *in-situ* transfer chamber system (made by Eiko Co., without air exposure from thin-film deposition chamber to TDS measurement chamber at ca. $10^{-8}$ Pa) | – (Sample is usually exposed to air before measurement.) |
|---|---|---|
| | Manipulators which pick-up sample from substrate carrier and set-up it on the TDS sample stage (made by Eiko Co.) | |
| Substrate for thin-film deposition (see FIG. 8 and Section VI) | Material | High-purity thin Si wafer (purity: 11 N, size: 1cm × 1cm × 100 μm in thickness, $H_2$ impurity content: $2.4 \times 10^{17}$ atoms/cm$^3$) | Not specified |
| | Pre-treatment before thin-film deposition | Thermal annealing at 800°C for 1h under UHV atmosphere ($H_2$ impurity content after annealing: $4.5 \times 10^{14}$ atoms/cm$^3$ | |

Figures

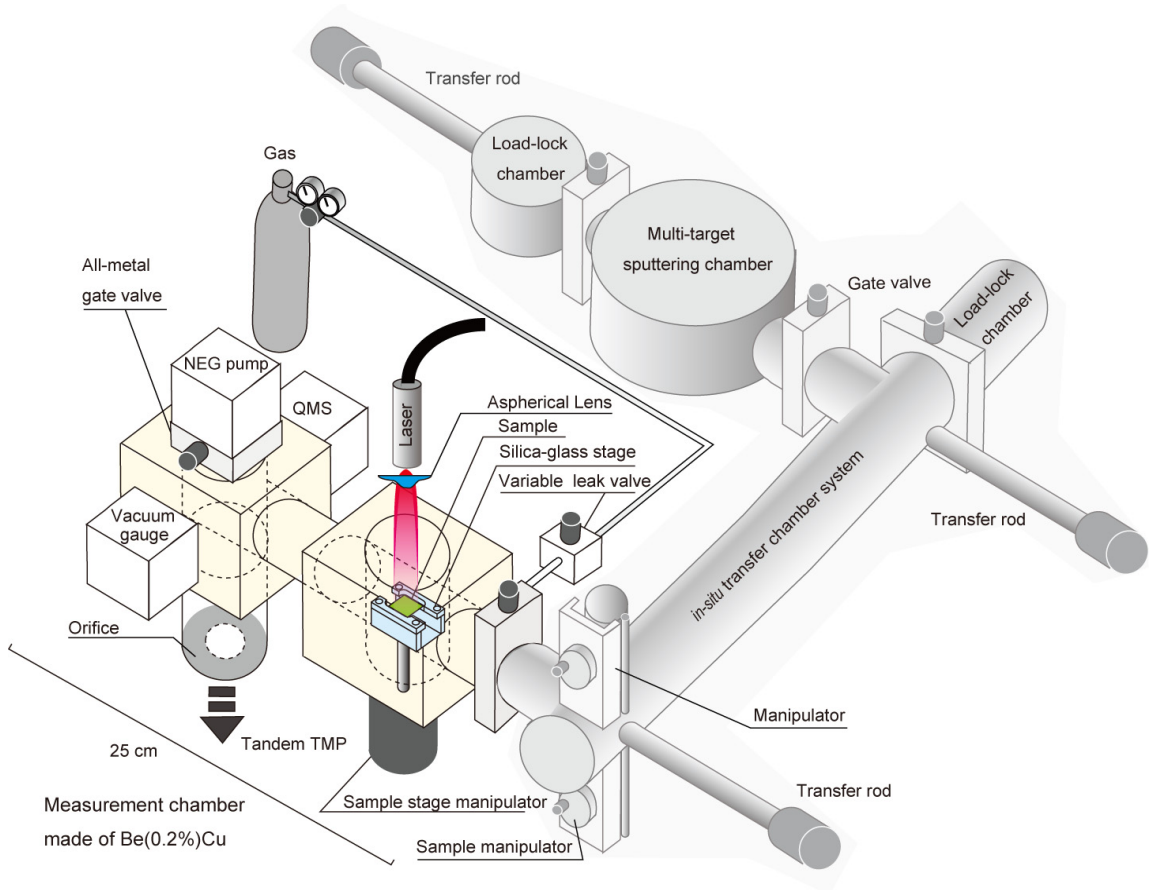

FIG. 1. Schematic illustration of the HHS-TDS system, to which an *in situ* transfer chamber, manipulators, and an rf magnetron sputtering deposition chamber are connected. The volume of the HHS-TDS measurement chamber, made of Be(0.2%)Cu, is as compact as 565 ml. The exhaust velocity is reduced to a low level with an orifice on top of a tandem TMP vacuum exhaust system. By employing this vacuum exhaust system, an UHV level of $9 \times 10^{-10}$ Pa as well as a low hydrogen residual partial pressure are achieved. To measure this UHV level, an ultralow outgas vacuum gauge (3B Gauge) is employed. Before thin-film deposition, a high-purity thin Si wafer substrate (size: 1 cm × 1 cm × 100 μm in thickness) is thermally annealed at 800 °C for 1 h under an UHV atmosphere in the multitarget rf magnetron sputtering thin-film deposition chamber, and then, the sample after film deposition is transferred under an UHV

atmosphere at approximately $10^{-8}$ Pa without air exposure to a point-contact high-purity silica-glass stage in the HHS-TDS measurement chamber using the *in situ* transfer chamber system and the manipulators to pick the sample up from the substrate carrier and place it on the sample stage. The thin-film sample on the stage is heated at a rate of 0.55 °C/s with an infrared laser diode heating system, which is precisely controlled by the dc input current to the laser diode. The laser light is irradiated through an aspherical lens from outside the measurement chamber. Simultaneously, thermally desorbed hydrogen ($m/z = 2$) is detected with an ultralow outgas QMS (WAT-Mass). For further improvement of the hydrogen residual partial pressure, a NEG pump (Magic NEG) is used directly before each TDS measurement. $H_2$ or $D_2$ gas is introduced from a side port of the measurement chamber using a variable leak valve only when the QMS signal is calibrated for quantitative analysis of hydrogen.

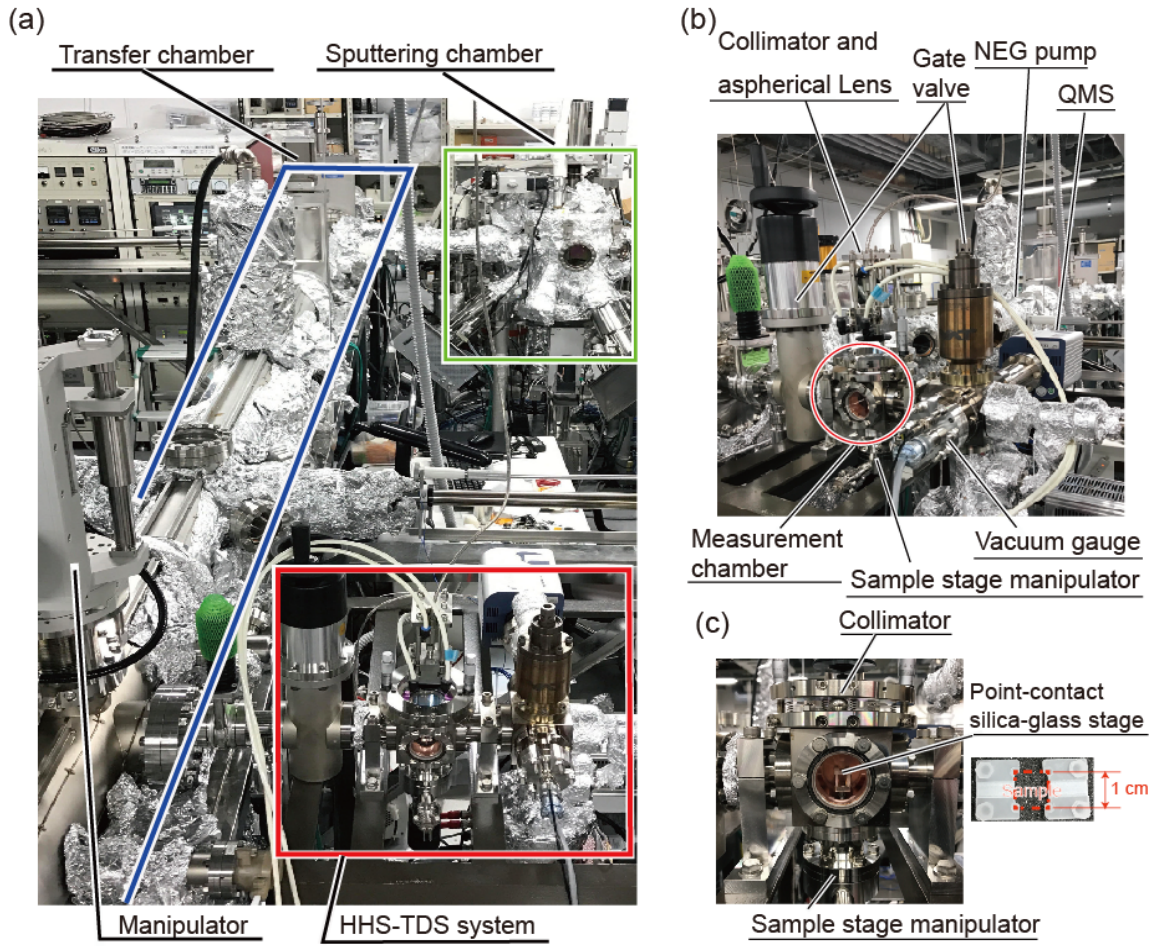

FIG. 2. Photographs of the HHS-TDS system. (a) Overview of the HHS-TDS system, with the *in situ* sample-transfer chamber system, a manipulator, and an rf magnetron sputtering thin-film deposition chamber, on a part (ca. 5 m × 5 m) of one experiment floor. All of these chambers are connected under an UHV condition of approximately $10^{-8}$ Pa. By employing this system, TDS measurements can be performed without exposing the thin-film samples to air. (b) HHS-TDS system. (c) Measurement chamber. The right photograph presents a top view of the point-contact silica-glass sample stage in the measurement chamber. The dotted square represents the sample position (lateral sample size: 1 cm × 1 cm).

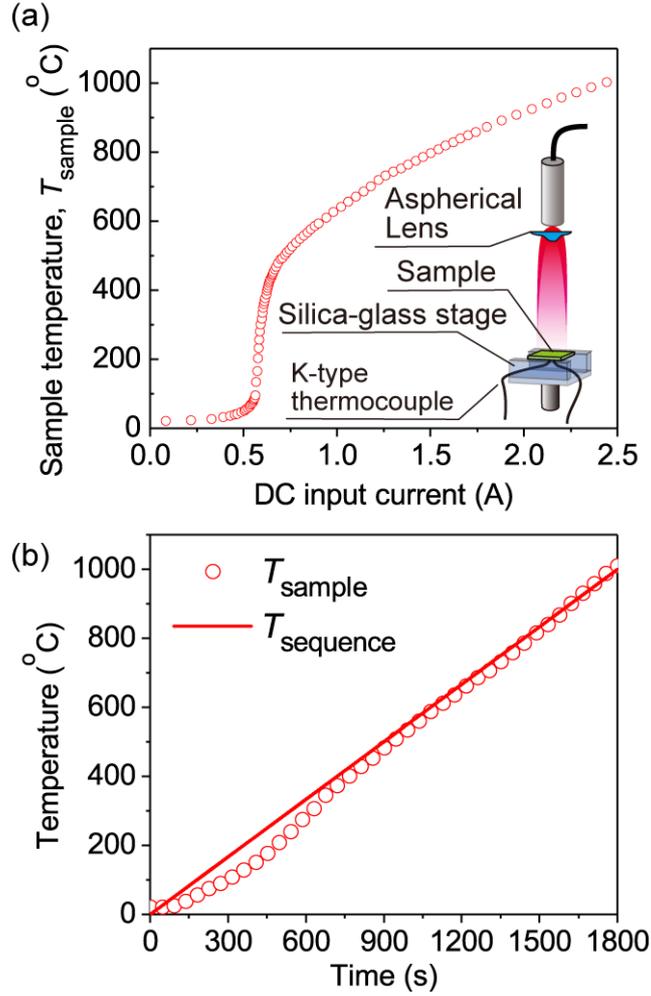

FIG. 3. Temperature calibration for the HHS-TDS system. (a) Relationship between $T_{sample}$ in the HHS-TDS and the dc input current to the laser diode heating system. The calibrated $T_{sample}$ is directly measured with a K-type thermocouple directly contacting the bottom of a Si substrate. The inset illustration shows the setup of this calibration experiment. (b) Relationship between the calibrated $T_{sample}$ and the temperature sequence at a heating rate of 0.55 °C/s ($T_{sequence}$).

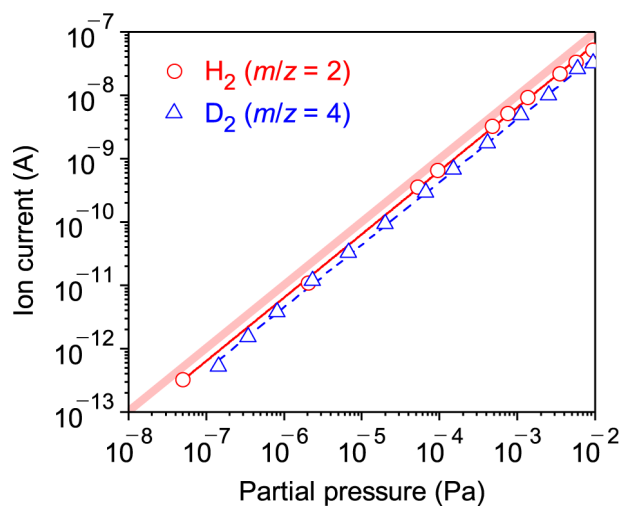

FIG. 4. Relationship between the ion current measured with the QMS (WAT-Mass) and the partial pressure of externally introduced H$_2$ (*m/z* = 2, circles) and D$_2$ (*m/z* = 4, triangles) gases measured with the vacuum gauge (3B gauge). The solid and dotted straight lines are the results obtained by least-squares fitting of the ion currents of the introduced H$_2$ and D$_2$ gases, respectively. The bold red straight line with a slope of 1 is shown for comparison.

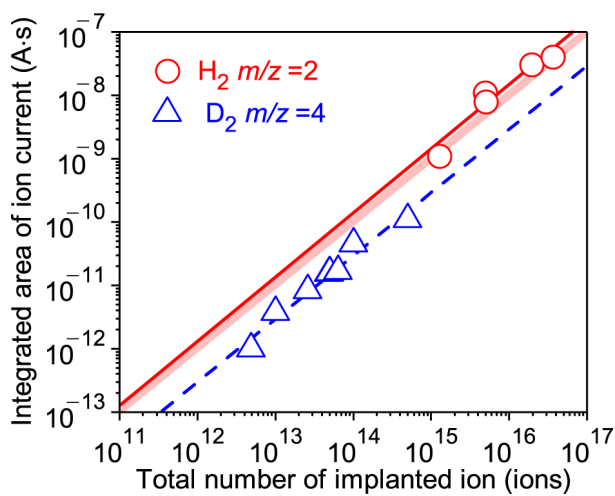

FIG. 5. Calibration result of the QMS signals for quantitative analysis of hydrogen concentration by HHS-TDS. The ion current of $H^+$- or $D^+$-ion-implanted standard Si wafer samples was measured up to 1000 °C at a heating rate of 0.55 °C/s. Integrated areas of the ion currents for $H_2$ $m/z$ = 2 (circles) and $D_2$ $m/z$ = 4 (triangles) are plotted against the total number of implanted $H^+$ or $D^+$ ions (i.e., dose amounts (ions/cm$^2$) × dose area (cm$^2$)) in the standard samples. The solid and dotted lines are the results obtained by least-squares fitting for $H_2$ and $D_2$, respectively. The bold red straight line with a slope of 1 is shown for comparison.

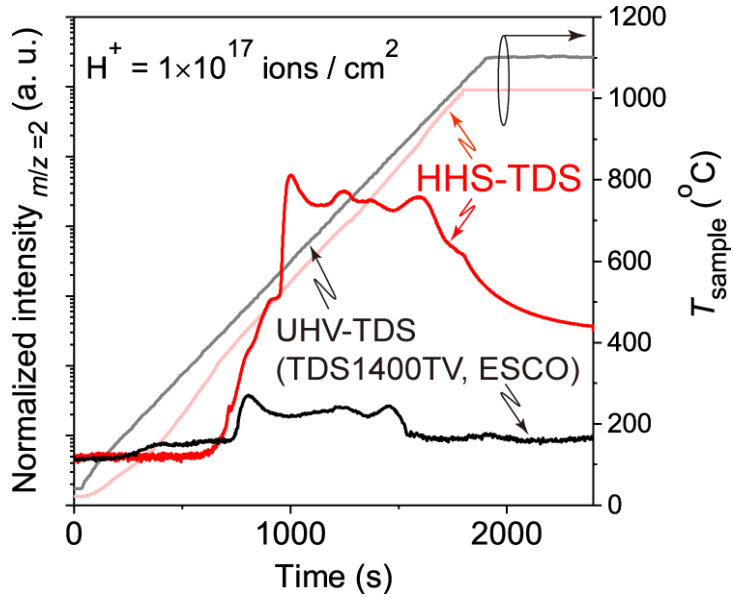

FIG. 6. Comparison of the QMS signal for $H_2$ ($m/z$ = 2) of the HHS-TDS system with that of a commercially available UHV-TDS system (TDS1400TV, ESCO Ltd.). A commercially available standard Si-wafer sample with dose amounts of implanted $H^+$ = $1 \times 10^{17}$ ions/cm$^2$ was employed for both measurements. Both the QMS signals were normalized at room temperature (i.e., measurement time = 0 s). The relationship between $T_{sample}$ and the measurement time for each TDS measurement is also shown using the right vertical axis. The hydrogen detection limits estimated from both the S/N ratios are $2.6 \times 10^{16}$ and $5.9 \times 10^{19}$ atoms/cm$^3$ for the HHS-TDS and UHV-TDS systems, respectively.

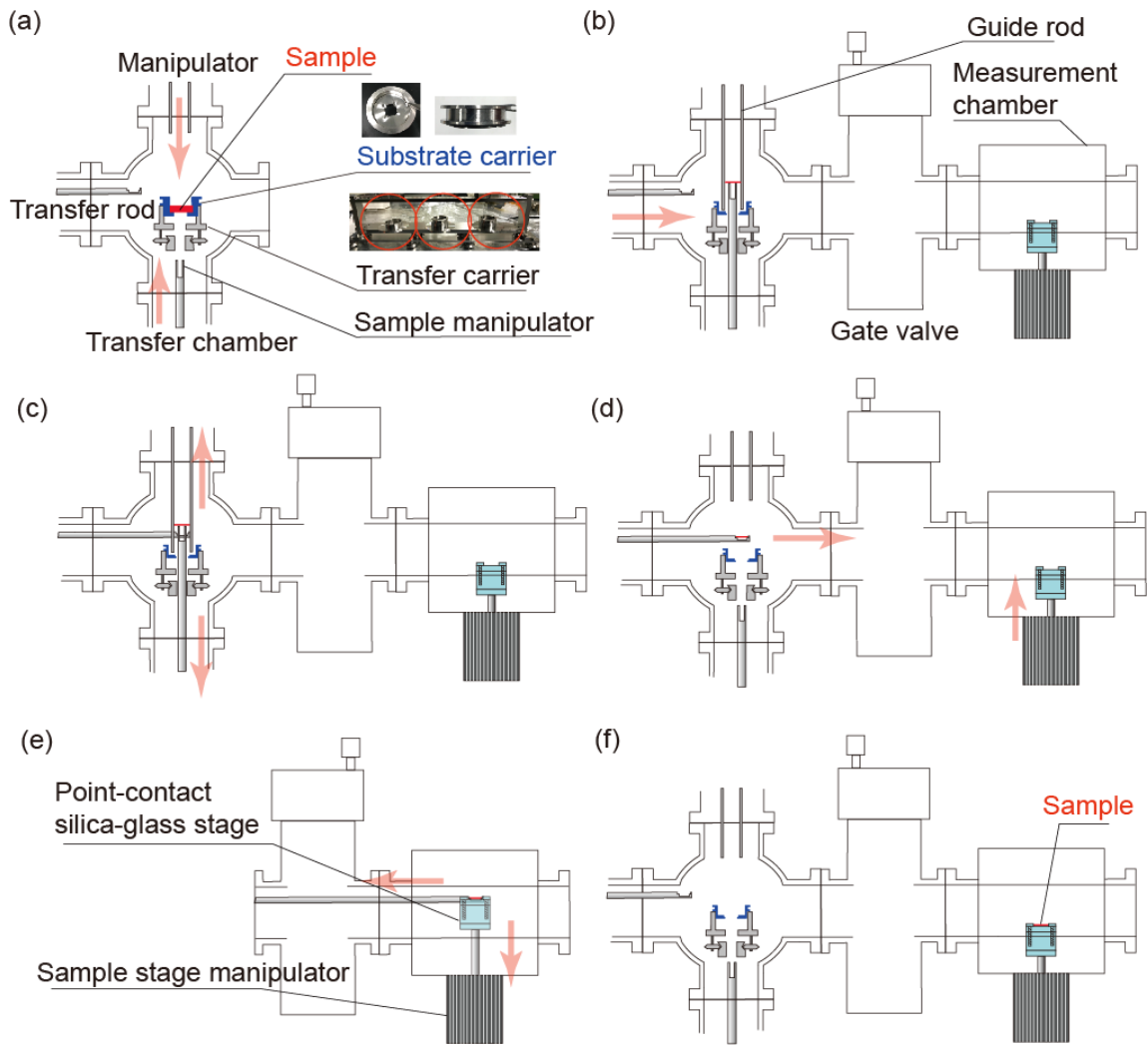

FIG. 7. Schematic illustrations of the transfer method of thin-film samples from the *in situ* transfer chamber system to the sample stage in the HHS-TDS measurement chamber using manipulators, which pick up the thin-film sample on the Si-wafer substrate from a substrate carrier and place it on the TDS sample stage. The transfer sequence is in the order of (a)→(b)→…→ and (f). Each action for the next transfer step is marked by arrows. (a) The thin-film sample is transferred from the rf sputtering deposition chamber to just beneath the manipulators near the TDS measurement chamber, see Figs. 1 and 2. The photographs (top) show a substrate carrier (left: top view, right: side view), which is composed of a plate for setting the substrate and a substrate carrier, and (bottom) a transfer carrier in the *in situ* transfer chamber system. The maximum number of substrate carriers we can carry using this transfer carrier is 3, as indicated by the circles. (b) A guide rod in the manipulator is loaded from the upper side to lock the plate and substrate carrier. Then, the thin-film sample is separated from the bottom side with the sample manipulator. (c, d) A transfer rod picks up the separated thin-film sample. (e, f) The sample is transferred to the TDS measurement chamber and placed on the silica-glass stage using the sample stage manipulator. (f) The final state directly before the TDS measurement.

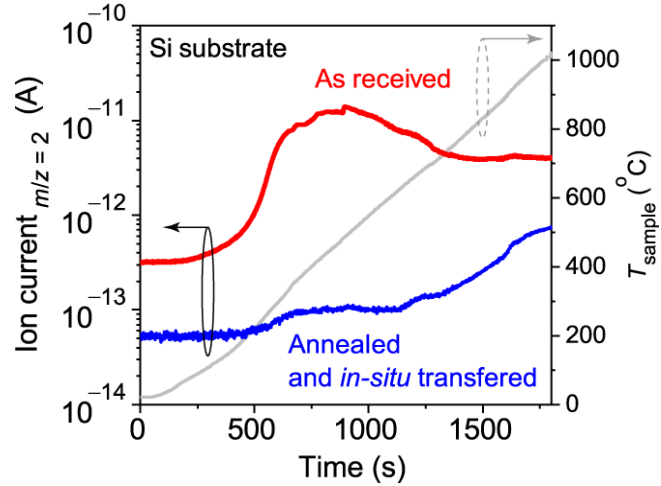

FIG. 8. HHS-TDS spectra for hydrogen ($m/z$ = 2) of as-received (red) and thermally annealed and *in situ* transferred (blue) Si substrates (purity: 11 N). The as-received Si substrate was first inserted into the load-lock chamber of the *in situ* transfer system from an air atmosphere and then transferred to the HHS-TDS measurement chamber without any pre-treatment. In contrast, the 'annealed' Si substrate was first transferred to the rf sputtering thin-film deposition chamber and then thermally annealed at 800 °C for 1 h in the deposition chamber. Next, the annealed substrate was *in situ* transferred to the HHS-TDS measurement chamber after being stored in the deposition chamber for 1 day without exposure to air. These results indicate that surface-adsorbed and impurity-hydrogen species in the Si-wafer substrate are removed to the background level of the QMS ($5 \times 10^{-14}$ A) by employing the pre-annealing step and *in situ* transfer. The estimated hydrogen concentration in the as-received and annealed Si substrates are $2.4 \times 10^{17}$ and $4.5 \times 10^{14}$ atoms/cm$^3$, respectively.

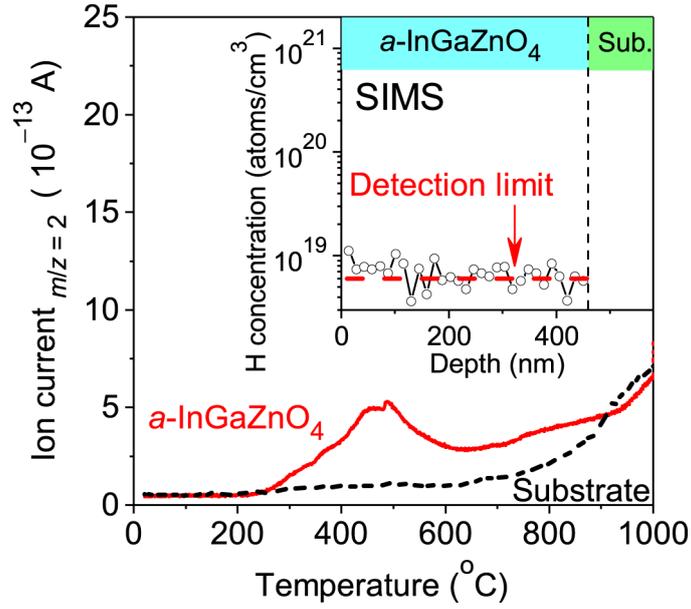

FIG. 9. Quantitative analysis of hydrogen concentration in a 1-μm-thick amorphous InGaZnO$_4$ (*a*-IGZO) thin film on a Si wafer substrate (Sub., lateral size: 1 cm × 1 cm) with a low hydrogen concentration using the HHS-TDS system. The Si substrate was pre-annealed at 800 °C for 1 h in the deposition chamber before deposition, and then, the as-deposited sample was transferred without exposure to air using the *in situ* transfer chamber system. To quantitatively analyze the hydrogen concentration, the TDS signal of the substrate (dotted line) was subtracted from that from the *a*-IGZO thin-film sample (solid line). This result indicates that the hydrogen concentration in the *a*-IGZO thin film was as low as $4.5 \times 10^{17}$ atoms/cm$^3$ and that the obtained maximum S/N ratio of the hydrogen ion current signal from the *a*-IGZO thin film was $1.4 \times 10^2$. The estimated hydrogen detection limit of the HHS-TDS system was as sensitive as $9.5 \times 10^{15}$ atoms/cm$^3$. The inset presents a depth profile of the SIMS measurement for the same sample for comparison, indicating that the hydrogen detection limit of the SIMS system is $6.6 \times 10^{18}$ atoms/cm$^3$.

Supplementary Material for 'Highly hydrogen-sensitive thermal desorption spectroscopy system for quantitative analysis of low hydrogen concentration (~1 × 10$^{16}$ atoms/cm$^3$) in thin-film samples'

Authors: Taku Hanna, Hidenori Hiramatsu, Isao Sakaguchi, and Hideo Hosono

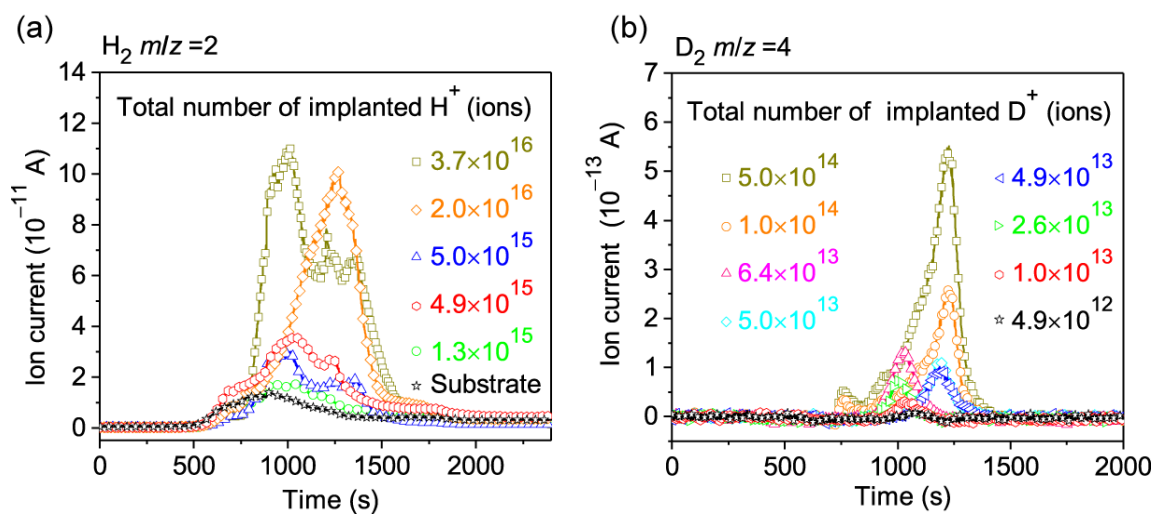

FIG. S1. Time dependence of the QMS ion current spectra for H$_2$ ($m/z$ = 2) and D$_2$ ($m/z$ = 4) of (a) H$^+$- and (b) D$^+$-implanted standard Si samples for the HHS-TDS system. All the spectra were obtained from room temperature to 1000 °C at a heating rate of 0.55 °C/s. After reaching 1000 °C, the temperature was maintained at 1000 °C during measurements (i.e., up to 2000 s) as seen in Fig. 6. The total number of implanted H$^+$ or D$^+$ ions for each standard sample was counted from the dose amounts of the implanted ions and dose area. All the implanted ions were completely desorbed after approximately 1800 s. The integrated areas of the ion currents are plotted against the total number of implanted H$^+$ or D$^+$ ions (i.e., dose amounts (ions/cm$^2$) × dosed area (cm$^2$)) in Fig. 5.

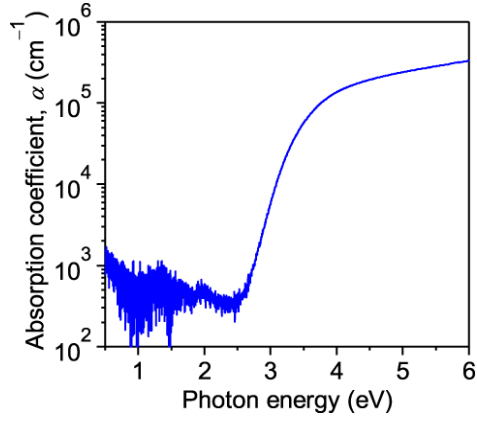

FIG. S2. Optical absorption coefficient ($\alpha$) spectrum of an *a*-IGZO thin film deposited using multitarget rf magnetron sputtering with an UHV back pressure of ~$10^{-8}$ Pa. An optically transparent silica-glass substrate was used instead of a Si substrate to measure the transmittance (*T*) and normal reflectance (*R*) spectra in the ultraviolet–visible–near-infrared wavelength region with a conventional spectrometer and to estimate $\alpha$ of the film using the relationship $\alpha = \ln[(1-R)/T]/d$, where *d* is the film thickness.